\pdfoutput=1

\documentclass[12pt,a4paper]{article}

\usepackage{ifthen} 
\newboolean{pdflatex}
\setboolean{pdflatex}{true} 

\newboolean{articletitles}
\setboolean{articletitles}{true} 

\newboolean{uprightparticles}
\setboolean{uprightparticles}{false} 

\def\paperauthors{LHCb collaboration} 
\def\papertitle{Search for \CP violation and observation of $P$ violation in \LbToppipipi decays} 
 
\def\papercopyright{\the\year\ CERN for the benefit of the LHCb collaboration} 
\def\paperlicence{CC-BY-4.0 licence}
\def\paperlicenceurl{https://creativecommons.org/licenses/by/4.0/}

\usepackage[top=1in, bottom=1.25in, left=1in, right=1in]{geometry}

\usepackage{microtype}
\usepackage{lineno}  
\usepackage{xspace} 
\usepackage{caption} 

\usepackage{graphicx}  
\usepackage{color}
\usepackage{colortbl}
\graphicspath{{./figs/}} 

\usepackage{amsmath} 
\usepackage{amssymb}
\usepackage{amsfonts}
\usepackage{upgreek} 

\newcommand*\patchAmsMathEnvironmentForLineno[1]{%
\expandafter\let\csname old#1\expandafter\endcsname\csname #1\endcsname
\expandafter\let\csname oldend#1\expandafter\endcsname\csname
end#1\endcsname
 \renewenvironment{#1}%
   {\linenomath\csname old#1\endcsname}%
   {\csname oldend#1\endcsname\endlinenomath}%
}
\newcommand*\patchBothAmsMathEnvironmentsForLineno[1]{%
  \patchAmsMathEnvironmentForLineno{#1}%
  \patchAmsMathEnvironmentForLineno{#1*}%
}
\AtBeginDocument{%
\patchBothAmsMathEnvironmentsForLineno{equation}%
\patchBothAmsMathEnvironmentsForLineno{align}%
\patchBothAmsMathEnvironmentsForLineno{flalign}%
\patchBothAmsMathEnvironmentsForLineno{alignat}%
\patchBothAmsMathEnvironmentsForLineno{gather}%
\patchBothAmsMathEnvironmentsForLineno{multline}%
\patchBothAmsMathEnvironmentsForLineno{eqnarray}%
}

\usepackage{hyperref}    
\usepackage[all]{hypcap} 

\def\lhcb {\mbox{LHCb}\xspace}

\def\MagUp {\mbox{\em Mag\kern -0.05em Up}\xspace}

\ifthenelse{\boolean{uprightparticles}}%
{

 \def\Ppi         {\ensuremath{\uppi}\xspace}

 \def\PDelta      {\ensuremath{\Delta}\xspace}                 
 \def\PXi      {\ensuremath{\Xi}\xspace}                 
 \def\PLambda      {\ensuremath{\Lambda}\xspace}                 
 \def\PSigma      {\ensuremath{\Sigma}\xspace}                 
 \def\POmega      {\ensuremath{\Omega}\xspace}                 
 \def\PUpsilon      {\ensuremath{\Upsilon}\xspace}                 
 
 \def\PB      {\ensuremath{\mathrm{B}}\xspace}                 
                  
 \def\PD      {\ensuremath{\mathrm{D}}\xspace}

 \def\PK      {\ensuremath{\mathrm{K}}\xspace}

 \def\Pb      {\ensuremath{\mathrm{b}}\xspace}                 
 \def\Pc      {\ensuremath{\mathrm{c}}\xspace}                 
 \def\Pd      {\ensuremath{\mathrm{d}}\xspace}

 \def\Pi      {\ensuremath{\mathrm{i}}\xspace}

 \def\Pp      {\ensuremath{\mathrm{p}}\xspace}

 \def\Pt      {\ensuremath{\mathrm{t}}\xspace}                 
 \def\Pu      {\ensuremath{\mathrm{u}}\xspace}

}
{

 \def\Ppi         {\ensuremath{\pi}\xspace}

 \mathchardef\PDelta="7101
 \mathchardef\PXi="7104
 \mathchardef\PLambda="7103
 \mathchardef\PSigma="7106
 \mathchardef\POmega="710A
 \mathchardef\PUpsilon="7107
                  
 \def\PB      {\ensuremath{B}\xspace}                 
                  
 \def\PD      {\ensuremath{D}\xspace}

 \def\PK      {\ensuremath{K}\xspace}

 \def\Pb      {\ensuremath{b}\xspace}                 
 \def\Pc      {\ensuremath{c}\xspace}                 
 \def\Pd      {\ensuremath{d}\xspace}

 \def\Pi      {\ensuremath{i}\xspace}

 \def\Pp      {\ensuremath{p}\xspace}

 \def\Pt      {\ensuremath{t}\xspace}                 
 \def\Pu      {\ensuremath{u}\xspace}

}

\makeatletter
\ifcase \@ptsize \relax
  \newcommand{\miniscule}{\@setfontsize\miniscule{4}{5}}
\or
  \newcommand{\miniscule}{\@setfontsize\miniscule{5}{6}}
\or
  \newcommand{\miniscule}{\@setfontsize\miniscule{5}{6}}
\fi
\makeatother

\DeclareRobustCommand{\optbar}[1]{\shortstack{{\miniscule (\rule[.5ex]{1.25em}{.18mm})}
  \\ [-.7ex] $#1$}}

\def\uquark    {{\ensuremath{\Pu}}\xspace}

\def\dquark    {{\ensuremath{\Pd}}\xspace}

\def\cquark    {{\ensuremath{\Pc}}\xspace}

\def\bquark    {{\ensuremath{\Pb}}\xspace}

\def\tquark    {{\ensuremath{\Pt}}\xspace}

\def\pion   {{\ensuremath{\Ppi}}\xspace}

\def\pip    {{\ensuremath{\pion^+}}\xspace}
\def\pim    {{\ensuremath{\pion^-}}\xspace}

\def\kaon    {{\ensuremath{\PK}}\xspace}
  \def\Kbar    {{\kern 0.2em\overline{\kern -0.2em \PK}{}}\xspace}

\def\KorKbar    {\kern 0.18em\optbar{\kern -0.18em K}{}\xspace}

\def\Kp      {{\ensuremath{\kaon^+}}\xspace}
\def\Km      {{\ensuremath{\kaon^-}}\xspace}

  \def\Dbar    {{\kern 0.2em\overline{\kern -0.2em \PD}{}}\xspace}
\def\D       {{\ensuremath{\PD}}\xspace}

\def\DorDbar    {\kern 0.18em\optbar{\kern -0.18em D}{}\xspace}

\def\B       {{\ensuremath{\PB}}\xspace}
\def\Bbar    {{\ensuremath{\kern 0.18em\overline{\kern -0.18em \PB}{}}}\xspace}

\def\BorBbar    {\kern 0.18em\optbar{\kern -0.18em B}{}\xspace}
\def\Bz      {{\ensuremath{\B^0}}\xspace}

  \def\Y#1S{\ensuremath{\PUpsilon{(#1S)}}\xspace}

\def\proton      {{\ensuremath{\Pp}}\xspace}
\def\antiproton  {{\ensuremath{\overline \proton}}\xspace}

\def\Deltares    {{\ensuremath{\PDelta}}\xspace}

\def\Lz          {{\ensuremath{\PLambda}}\xspace}
\def\Lbar        {{\ensuremath{\kern 0.1em\overline{\kern -0.1em\PLambda}}}\xspace}
\def\LorLbar    {\kern 0.18em\optbar{\kern -0.18em \PLambda}{}\xspace}

\def\Lb      {{\ensuremath{\Lz^0_\bquark}}\xspace}
\def\Lbbar   {{\ensuremath{\Lbar{}^0_\bquark}}\xspace}
\def\Lc      {{\ensuremath{\Lz^+_\cquark}}\xspace}

\newcommand{\decay}[2]{\ensuremath{#1\!\to #2}\xspace}         

\def\to                 {\ensuremath{\rightarrow}\xspace}

\def\CP                {{\ensuremath{C\!P}}\xspace}

\def\Vub      {{\ensuremath{V_{\uquark\bquark}^{\phantom{*}}}}\xspace}

\def\Vtb      {{\ensuremath{V_{\tquark\bquark}^{\phantom{*}}}}\xspace}
\def\Vudconj  {{\ensuremath{V_{\uquark\dquark}^*}}\xspace}

\def\Vtdconj  {{\ensuremath{V_{\tquark\dquark}^*}}\xspace}

\newcommand{\AT}{{\ensuremath{A_{\widehat{T}}}}\xspace}
\newcommand{\ATbar}{{\ensuremath{\kern 0.1em\overline{\kern -0.1em A}_{\widehat{T}}}}\xspace}
\newcommand{\aCPTodd}{{\ensuremath{a_\CP^{\widehat{T}\text{-odd}}}}\xspace}
\newcommand{\aPTodd}{{\ensuremath{a_P^{\widehat{T}\text{-odd}}}}\xspace}

\newcommand{\CT}{{\ensuremath{C_{\widehat{T}}}}\xspace}
\newcommand{\CTbar}{{\ensuremath{\kern 0.1em\overline{\kern -0.1em C}_{\widehat{T}}}}\xspace}

\def\LbTopKpipi   {\decay{\Lb}{\proton\Km\pip\pim}}

\def\LbToppipipi  {\decay{\Lb}{\proton\pim\pip\pim}}

\def\LbToLcpiCS   {\decay{\Lb}{\Lc(\decay{}{\proton\Km\pip})\pim}}

\def\That       {\ensuremath{\widehat{T}}\xspace}

\def\C#1      {\ensuremath{\mathcal{C}_{#1}}\xspace}                       
\def\Cp#1     {\ensuremath{\mathcal{C}_{#1}^{'}}\xspace}                    
\def\Ceff#1   {\ensuremath{\mathcal{C}_{#1}^{\mathrm{(eff)}}}\xspace}        
\def\Cpeff#1  {\ensuremath{\mathcal{C}_{#1}^{'\mathrm{(eff)}}}\xspace}       
\def\Ope#1    {\ensuremath{\mathcal{O}_{#1}}\xspace}                       
\def\Opep#1   {\ensuremath{\mathcal{O}_{#1}^{'}}\xspace}                    

\newcommand{\aunit}[1]{\ensuremath{\text{\,#1}}}       

\newcommand{\tev}{\aunit{Te\kern -0.1em V}\xspace}
\newcommand{\gev}{\aunit{Ge\kern -0.1em V}\xspace}
\newcommand{\mev}{\aunit{Me\kern -0.1em V}\xspace}
\newcommand{\kev}{\aunit{ke\kern -0.1em V}\xspace}
\newcommand{\ev}{\aunit{e\kern -0.1em V}\xspace}
\newcommand{\mevc}{\ensuremath{\aunit{Me\kern -0.1em V\!/}c}\xspace}
\newcommand{\gevc}{\ensuremath{\aunit{Ge\kern -0.1em V\!/}c}\xspace}
\newcommand{\mevcc}{\ensuremath{\aunit{Me\kern -0.1em V\!/}c^2}\xspace}
\newcommand{\gevcc}{\ensuremath{\aunit{Ge\kern -0.1em V\!/}c^2}\xspace}

\def\invfb   {\ensuremath{\mbox{\,fb}^{-1}}\xspace}

\newcommand{\chisq}{\ensuremath{\chi^2}\xspace}

\def\gsim{{~\raise.15em\hbox{$>$}\kern-.85em
          \lower.35em\hbox{$\sim$}~}\xspace}
\def\lsim{{~\raise.15em\hbox{$<$}\kern-.85em
          \lower.35em\hbox{$\sim$}~}\xspace}

\def\evtgen     {\mbox{\textsc{EvtGen}}\xspace}

\def\geant      {\mbox{\textsc{Geant4}}\xspace}

\def\photos     {\mbox{\textsc{Photos}}\xspace}

\def\pythia     {\mbox{\textsc{Pythia}}\xspace}

\def\tell1  {TELL1\xspace}
\def\ukl1   {UKL1\xspace}

\newcommand{\eg}{\mbox{\itshape e.g.}\xspace}

\usepackage{cite} 
\usepackage{mciteplus}

\usepackage[normalem]{ulem}

\usepackage{booktabs}
\usepackage{multirow}

\usepackage{placeins}
\usepackage{longtable} 

\begin{document}


\renewcommand{\thefootnote}{\fnsymbol{footnote}}
\setcounter{footnote}{1}


\begin{titlepage}
\pagenumbering{roman}

\vspace*{-1.5cm}
\centerline{\large EUROPEAN ORGANIZATION FOR NUCLEAR RESEARCH (CERN)}
\vspace*{1.5cm}
\noindent
\begin{tabular*}{\linewidth}{lc@{\extracolsep{\fill}}r@{\extracolsep{0pt}}}
\ifthenelse{\boolean{pdflatex}}
{\vspace*{-1.5cm}\mbox{\!\!\!\includegraphics[width=.14\textwidth]{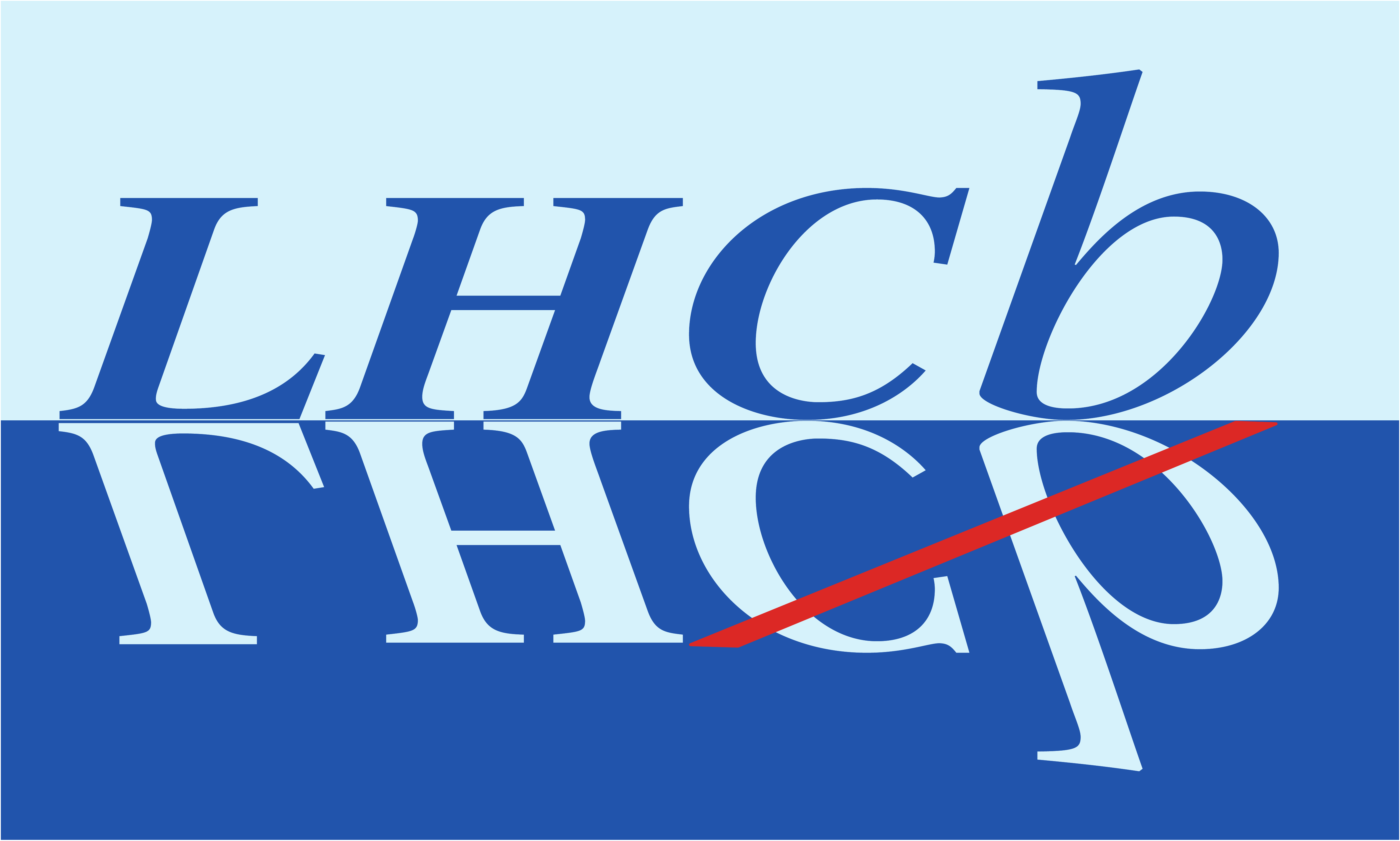}} & &}%
{\vspace*{-1.2cm}\mbox{\!\!\!\includegraphics[width=.12\textwidth]{figs/lhcb-logo.eps}} & &}%
\\
 & & CERN-EP-2019-256 \\  
 & & LHCb-PAPER-2019-028 \\  
 & & \today \\ 
 & & \\
\end{tabular*}

\vspace*{3.0cm}

{\normalfont\bfseries\boldmath\huge
\begin{center}
  \papertitle 
\end{center}
}

\vspace*{1.5cm}

\begin{center}
\paperauthors\footnote{Authors are listed at the end of this Letter.}
\end{center}

\vspace{\fill}

\begin{abstract}
  \noindent
  A search for \CP violation in the \LbToppipipi decay is performed using LHCb data corresponding to an integrated luminosity of 6.6\invfb collected in $pp$ collisions at centre-of-mass energies of 7, 8 and 13\tev. 
The analysis uses both triple product asymmetries and the unbinned energy test method. 
The highest significances of \CP asymmetry are 2.9 standard deviations from triple product asymmetries and 3.0 standard deviations for the energy test method. Once the global $p$-value is considered, all results are consistent with  no \CP violation. 
Parity violation is observed at a significance of  $5.5$ standard deviations for the triple product asymmetry method and $5.3$ standard  deviations for the energy test method. The reported deviations are given in regions of phase space.
\end{abstract}

\vspace*{1.5cm}

\begin{center}
  Published in Phys. Rev. D 102 (2020) 051101
\end{center}

\vspace{\fill}

{\footnotesize 
\centerline{\copyright~\papercopyright. \href{\paperlicenceurl}{\paperlicence}.}}
\vspace*{2mm}

\end{titlepage}


\newpage
\setcounter{page}{2}
\mbox{~}

\cleardoublepage

\renewcommand{\thefootnote}{\arabic{footnote}}
\setcounter{footnote}{0}


\pagestyle{plain} 
\setcounter{page}{1}
\pagenumbering{arabic}


The violation of \CP symmetry, where $C$ and $P$ are the charge-conjugation and parity operators, is a well-established phenomenon in the decays of \kaon and \B
mesons~\cite{PhysRevLett.13.138,Aubert:2001sp,Abe:2001xe}. Recently, it has also been observed in the decays of \D mesons by the LHCb collaboration~\cite{LHCb-PAPER-2019-006}.
However, 
\CP violation has yet to be established in baryonic decays, although first evidence was recently found~\cite{LHCb-PAPER-2016-030}. 
Such decays offer a novel environment to probe the mechanism for quark-flavour mixing and for \CP violation, which is regulated by the Cabibbo-Kobayashi-Maskawa (CKM) matrix in the Standard Model (SM)~\cite{PhysRevLett.10.531,Kobayashi:1973fv}.

In this Letter searches for \CP and $P$ violation with \LbToppipipi decays are reported. Throughout, the inclusion of charge-conjugate processes is implied, unless otherwise indicated. This decay is  mediated mainly by tree and loop processes of similar magnitudes, 
proportional to the product of the CKM matrix elements \Vub\Vudconj and \Vtb\Vtdconj, respectively. This allows for significant interference effects with a  relative weak phase $\alpha$ of the Unitary Triangle between the amplitudes.  
If matter and antimatter exhibit different effects, \CP violation manifests as either global asymmetries in decay rates, or as local asymmetries within the phase space. The \LbToppipipi decay is particularly well suited for \CP-violation searches~\cite{Gronau:2015gha} due to a rich resonant structure in the decay. The dominant contributions proceed through the $\decay{N^{*+}}{\Deltares^{++} (1234)\pim}$ (referred as $\Deltares^{++}$ hereinafter), $\decay{\Deltares^{++}}{\proton\pip}$,  $a_{1}^{-}(1260)\to\rho^{0}(770)\pim$ and $\rho^{0}(770)\to\pip\pim$ decays, where the proton excited states are indicated as $N^{*+}$. The searches for \CP violation are performed by separating the $P$-odd and $P$-even contributions~\cite{Durieux:2016nqr}, as discussed below. 
In these studies, a large control sample of Cabibbo-favored \LbToLcpiCS decays is used, where no \CP violation is expected, to assess potential experimental biases and systematic effects.

The \lhcb collaboration has previously studied the  \LbToppipipi decay and found evidence for \CP violation with a  significance of 3.3 standard deviations including systematic uncertainties~\cite{LHCb-PAPER-2016-030}. This Letter supersedes the previous results using $pp$ collision data corresponding to an integrated luminosity of 6.6\invfb collected from 2011 to 2017 at centre-of-mass energies of 7, 8 and 13\tev that represents a four times larger sample in signal yield.

The LHCb 
detector~\cite{Alves:2008zz,LHCb-DP-2014-002} is a
single-arm forward spectrometer covering the pseudorapidity range $2 < \eta < 5$, designed for
the study of particles containing \bquark\ or \cquark\ quarks. The detector elements that are particularly
relevant to this analysis are: a silicon-strip vertex detector surrounding the $pp$ interaction
region that allows \bquark\ hadrons to be identified from their characteristically long
flight distance; a tracking system that provides a measurement of the momentum, $p$, of charged
particles; and two ring-imaging Cherenkov detectors that are able to discriminate between
different species of charged hadrons.
Simulation is required to model the effects of the detector acceptance and the selection requirements.
  The $pp$ collisions are generated using
\pythia~\cite{Sjostrand:2006za,*Sjostrand:2007gs} with a specific \lhcb
configuration~\cite{LHCb-PROC-2010-056}, and neither \CP- nor $P$-violating effects are present in the signal channel.  Decays of unstable particles
are described by \evtgen~\cite{Lange:2001uf}, in which final-state
radiation is generated using \photos~\cite{Golonka:2005pn}. The
interaction of the generated particles with the detector, and its response,
are implemented using the \geant
toolkit~\cite{Allison:2006ve, *Agostinelli:2002hh} as described in
Ref.~\cite{LHCb-PROC-2011-006}.

The analysis searches for \CP and $P$ violation by measuring triple product asymmetries (TPA) and by exploiting the unbinned energy test method~\cite{ASLAN00949650410001661440,ASLAN2005626,Williams:2010vh,Williams:2011cd,Parkes:2016yie,Barter:2018xbc,Gillam:2018ufy}. In the TPA analysis, both local and integrated asymmetries are considered. The analysis also benefits from additional studies of amplitude models~\cite{Durieux:2015zwa,Durieux:2016nqr} to maximise the sensitivity. The energy test method is designed to look for localized differences in the phase space between two samples. The \Lb polarization has been measured to be compatible with zero in a previous LHCb analysis~\cite{Aaij:2013oxa} and is neglected in these measurements.

The scalar triple products are defined as $\CT\equiv\vec{p}_\proton\cdot\left(\vec{p}_{\pi_\text{fast}^-} \times \vec{p}_{\pi^{+}}\right)$ and 
\mbox{$\CTbar\equiv\vec{p}_\antiproton\cdot\left(\vec{p}_{\pi_\text{fast}^+} \times \vec{p}_{\pi^{-}}\right)$}, for \Lb and \Lbbar respectively.
Hereinafter $\pi_\text{fast}^-$ ($\pi_\text{slow}^-$) refers to the faster (slower) of two negative pions in the \Lb rest frame. Following these definitions, four statistically independent subsamples are considered, labeled with \textit{I} for $\CT > 0$, \textit{II} for $\CT < 0$, \textit{III} for $-\CTbar > 0$ and \textit{IV} for $-\CTbar < 0$. Samples \textit{I} and \textit{III} are related by a \CP transformation, as are samples \textit{II} and \textit{IV}. Samples \textit{I} and \textit{II} are related by a $P$ transformation, as are samples \textit{III} and \textit{IV}. Both \CP- and $P$-violating effects appear as differences between the triple product observables related by \CP and $P$ transformations. 
The $\That$ operator reverses momentum and spin three-vectors~\cite{Sachs,Branco:1999fs}. The quantities \CT and \CTbar are odd under this operator. This enables  studies of the $P$-odd \CP violation, which occurs via interference of the \That-even and \That-odd amplitudes with different \CP-odd (`weak') phases~\cite{Sachs,Branco:1999fs,Durieux:2015zwa,Durieux:2016nqr}.

The TPA are defined as
\begin{equation}
\AT = \frac{N(\CT>0)-N(\CT<0)}{N(\CT>0)+N(\CT<0)}, \ATbar = \frac{\overline{N}(-\CTbar>0)-\overline{N}(-\CTbar<0)}{\overline{N}(-\CTbar>0)+\overline{N}(-\CTbar<0)}, 
\end{equation}
where $N$ and $\overline{N}$ are the yields of \Lb and \Lbbar decays, respectively. 
The \CP- and $P$-violating asymmetries are then defined as 
\begin{equation}
\aCPTodd = \frac{1}{2}\left(\AT- \ATbar\right), \, \aPTodd = \frac{1}{2}\left(\AT +\ATbar\right).
\end{equation}

Two types of asymmetries are determined from data. The first are localized in the phase space in order to enhance sensitivity to local effects and the second are integrated over the whole phase space. By construction, such asymmetries are largely insensitive to particle-antiparticle
production and detector-induced asymmetries~\cite{LHCB-PAPER-2014-046}. 

The previous LHCb result ~\cite{LHCb-PAPER-2016-030} showed evidence for a dependence of the \CP asymmetry as a function of $\lvert\Phi\rvert$, the absolute value of the angle between the planes defined by the $\proton\pi_\text{fast}^-$ and $\pip\pi_\text{slow}^-$ systems in the \Lb rest frame. In the present analysis a binning scheme, labeled \textit{A}, is considered, based on the results of an approximate amplitude analysis performed on \LbToppipipi decays. The binning scheme consists in dividing the data sample into 16 subsamples to explore the distribution of the polar and azimuthal angles of the proton ($\Deltares^{++}$) in the $\Deltares^{++}$ ($N^{*+}$) rest frame. A detailed description can be found in Appendix~\ref{app:definition binning A}. A second binning scheme, labeled \textit{B}, is used to probe the asymmetries as a function of $\lvert\Phi\rvert$,  dividing the data sample into ten subsamples uniformly distributed in the range $[0,\pi]$. 
The invariant-mass regions $m(\proton\pip\pi_{\text{slow}}^-) > 2.8\gevcc$ (samples $A_1,B_1$), dominated by the $a_1$ resonance, and $m(\proton\pip\pi_{\text{slow}}^-) < 2.8\gevcc$ (samples $A_2,B_2$), dominated by the $N^{*+}$ decay, are studied separately.
The compatibility of the measured asymmetries with \CP and $P$ conservation is checked by means of a \chisq test taking into account statistical and systematic effects.

The energy test is a model-independent unbinned test sensitive to local differences between two samples, as might arise from \CP violation. It can provide superior discriminating power between different samples than traditional \chisq tests~\cite{Williams:2011cd,Parkes:2016yie}. The test is performed through the calculation of a test statistic
\begin{equation}
T \equiv \frac{1}{2n(n-1)}\sum_{i\neq j}^{n}\psi_{ij}
+ \frac{1}{2\overline{n}(\overline{n}-1)}\sum_{i\neq j}^{\overline{n}}\psi_{ij}
- \frac{1}{n\overline{n}} \sum_{i=1}^{n}\sum_{j=1}^{\overline{n}}\psi_{ij} ,
\label{eqn:T}
\end{equation}
where there are $n$ ($\overline{n}$) candidates in the first (second) sample. The first (second) term sums over pairs of candidates drawn from the first (second) sample and the final term sums over pairs with one candidate drawn from each sample. Each pair of candidates $ij$ is assigned a weight  $\psi_{ij} = e^{-d^2_{ij}/2\delta^2}$, where $d_{ij}$ is their Euclidean distance in phase space, while the tunable parameter $\delta$ determines the distance scale probed using the energy test. The phase space is defined using the squared masses $m^{2}(p\pi^{+})$,
$m^{2}(\pi^{+}\pi^{-}_\text{slow})$, 
$m^{2}(p\pi^{+}\pi^{-}_\text{slow})$,
$m^{2}(\pi^{+}\pi^{-}_\text{slow}\pi^{-}_\text{fast})$
and $m^{2}(p\pi^{-}_\text{slow})$.  
 The value of  $T$ is large when there are significant localized  differences between samples and  has an expectation of zero when there are no differences. The distribution of $T$ under the hypothesis of no sample differences, and the assignment of $p$-values, are determined using a permutation method~\cite{Williams:2011cd,Barter:2018xbc}.

Similarly to the TPA method, the comparison of subsamples \textit{I} and \textit{IV} to subsamples \textit{II} and \textit{III} allows for a $P$-odd and \CP-odd test; the  comparison of subsamples \textit{I} and \textit{II} to subsamples \textit{III} and  \textit{IV} for a $P$-even and \CP-odd test. 
The $P$ violation is also tested by comparing the combination of subsamples \textit{I} and \textit{III} with the combination of subsamples \textit{II} and \textit{IV}. This provides three test configurations described in detail in Ref. \cite{Parkes:2016yie} and illustrated in figures therein.
The length scale at which \CP violation might appear is not known. Therefore three different scales are probed in each configuration, chosen following Refs.~\cite{Williams:2011cd,Parkes:2016yie} as $\delta = 1.6\text{ GeV}^{2}/c^{4}$, $2.7\text{ GeV}^{2}/c^{4}$ and $13\text{ GeV}^{2}/c^{4}$. The sensitivity of the chosen scales was confirmed using simulated events.
For each of the three test configurations all three scales are probed, such that nine tests are made overall: six tests for effects arising from \CP violation (three probing $P$-even \CP violation and three $P$-odd \CP violation) and three tests for effects arising from $P$ violation.

\label{sec:selection}

\begin{figure}[t]
\centering
\includegraphics[width=0.49\textwidth]{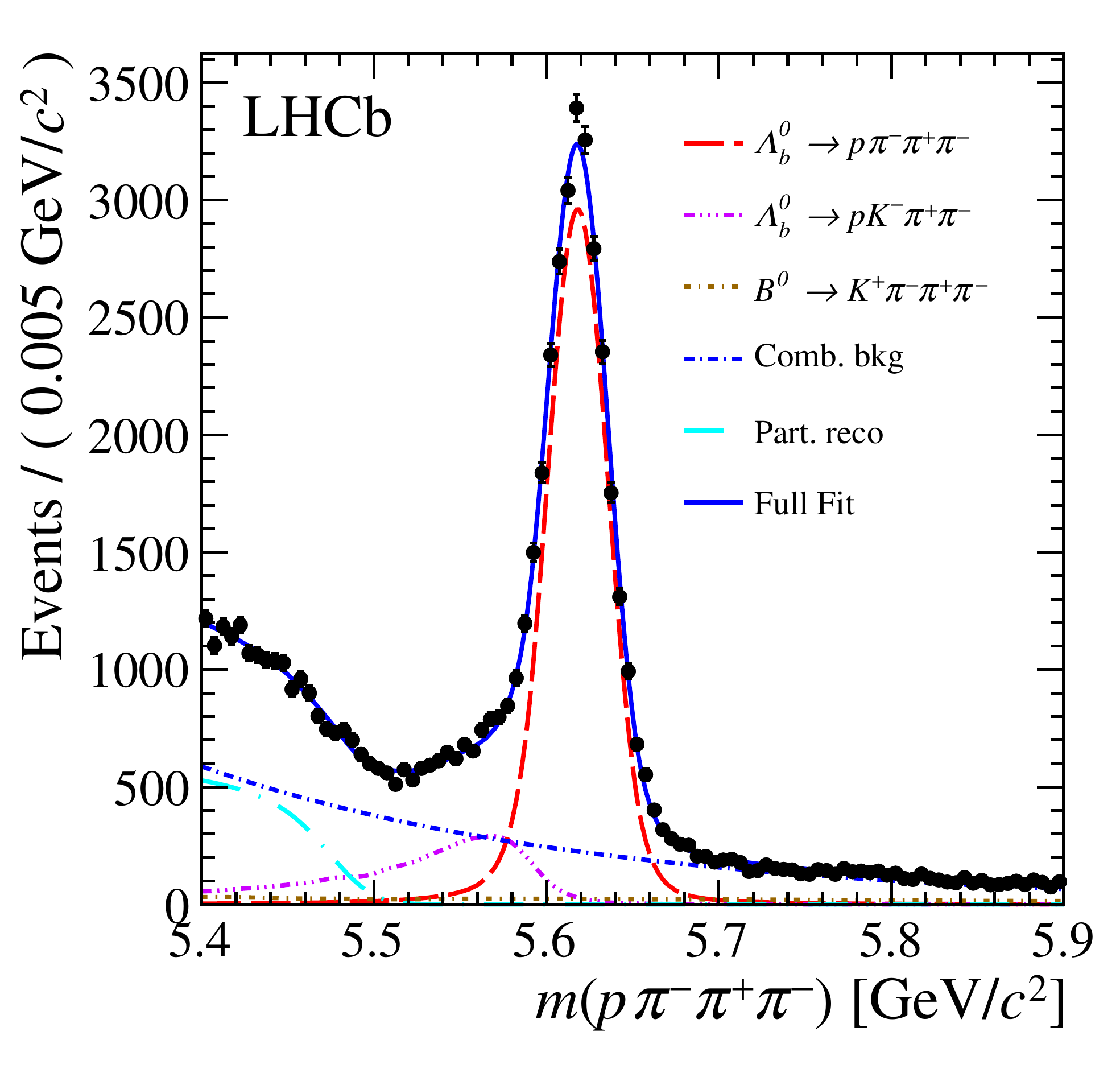}
\caption{\label{fig:p3pi_massfit} 
Invariant-mass distribution for \LbToppipipi candidates with the result of the fit overlaid. The solid and dotted lines describe the projections of the fit results for various components as listed in the legend.
}
\end{figure}

The candidate \LbToppipipi decays are formed by combining tracks with transverse (total) momentum greater than 250\mevc (1.5\gevc) identified as protons and pions that originate from a common vertex displaced from the primary vertex. A cut on the invariant-mass $m(\proton\Km\pip) \in [2.26,2.30]\gevcc$ is applied to select \LbToLcpiCS decay candidates used as a control sample.
A boosted decision tree classifier~\cite{Breiman} (BDT), independently optimised for different center-of-mass energies, is constructed from
a set of kinematic variables that discriminate between signal
and background. 
The result of an unbinned extended maximum-likelihood fit to the  invariant-mass distribution, $m(\proton\pim\pip\pim)$, is shown in Fig.~\ref{fig:p3pi_massfit} for the dataset integrated over the phase space.
The invariant-mass distribution of the signal is modelled by a Gaussian function core with power-law tails~\cite{Skwarnicki:1986xj}, with the mean and width of the Gaussian function determined from the fit to data. All other parameters of the signal fit model are taken from simulation except for the yields.
The combinatorial background is parameterised with an exponential function where the parameters are left free to vary in the fits.
Partially reconstructed \Lb decays, as for example $\decay{\Lb}{p\pim\pip\pim\pi^0}$, are described by an ARGUS function~\cite{Albrecht:1990am} convolved with a Gaussian function to account for resolution effects.
The shapes of backgrounds from other \bquark-hadron decays due to incorrectly identified particles,
\eg kaons identified as pions or protons identified as kaons,
are modelled using simulated events.
These consist mainly of \LbTopKpipi and \decay{\Bz}{\Kp\pim\pip\pim} decays. Their yields are obtained from fits to data where the invariant-mass distributions are reconstructed under the appropriate mass hypotheses and then fixed in the baseline fits. 
The signal yields for the \LbToppipipi decay and the  \LbToLcpiCS control sample are $27\,600\pm200$ and $434\,500\pm800$, respectively.  Fits in bins of phase space are also performed to determine asymmetries \AT and \ATbar in each region, assigning signal candidates to four categories according to \Lb or \Lbbar flavour and sign of \CT or \CTbar. The asymmetries \AT and \ATbar are found to be uncorrelated. 
Corresponding asymmetries for each of the background components are also determined in the fit;
they are found to be consistent with zero, and do not lead to significant systematic uncertainties in the signal asymmetries.
Artificial asymmetries are generated for signal events using a parameterised simulated sample, and used to perform checks of the sensitivity of the methods applied. When $P$-odd \CP violation is injected via the $N^*$ resonances in such studies, both the triple product asymmetry method and the energy test are able to provide a clear rejection of the no-\CP violation hypothesis. When $P$-even \CP violation is injected in the simulated samples via the $a_1$ resonance, the energy test is also able to observe this effect.

For the energy test, \Lb candidates are selected in  a window corresponding to 2.5 standard deviations of the Gaussian function around the known \Lb mass~\cite{PDG2018}, which optimises  the sensitivity to \CP violation. The background component with this selection is small and does not affect the analysis. 

The reconstruction efficiency for signal candidates with $\CT>0$ is consistent with that for candidates with $\CT<0$. This indicates that the detector and the reconstruction algorithms do not bias the measurements. This is confirmed using the control sample and a large sample of simulated events. 
The same check is performed for the \CTbar observable. As a general cross-check, the \CP asymmetry is measured in the control sample and  found to be compatible with zero, \mbox{$\aCPTodd(\Lc\pim) = (+0.04\pm0.16)\%$}. 

The main sources of systematic uncertainties in the TPA analysis are selection criteria, reconstruction and detector acceptance. They are evaluated using the control sample.
In the TPA analysis, a systematic uncertainty of $0.16\%$ is assigned for the integrated measurements, while uncertainties in the range  $(0.6\text{--}2.5)\%$ are assigned for local measurements.
The systematic uncertainty arising from the experimental resolution of the triple products \CT and \CTbar, which could introduce a migration of candidates between bins, is estimated from simulation. 
The difference between the reconstructed and generated asymmetries, $0.01\%$, is taken as a systematic uncertainty in the TPA analysis.
To assess the systematic uncertainty associated with the fit model, an alternative is used to compare the results measured on pseudoexperiments with respect to the baseline model. A value of $0.06\%$ ($0.08\%$) for \aCPTodd/\aPTodd (\AT/\ATbar) is assigned as systematic uncertainty. 
No significant differences are observed comparing results from different running conditions, 
trigger requirements and selection criteria. 

Several studies are  made to confirm the reliability of the energy test method. The method is insensitive  to global asymmetries, and so is not affected by differences between \Lb and \Lbbar production rates. However, local asymmetries due to detector effects may yield significant results that would lead to an incorrect conclusion. The potential presence of such effects is studied using the control sample. 
No evidence is found for any local asymmetry.

Contributions from background decays are considered, in case they contain
localized asymmetries not related to \CP violation. A high-mass selection is applied (\mbox{$5.75<m(p\pim\pip\pim)<6.10 \gevcc$}) to identify candidates predominantly produced by random combinations of particles. No significant effect is found in the six configurations of the energy test probing the \CP-conserving hypothesis. Moreover, a small independent sample of the dominant peaking background (\LbTopKpipi) is selected using the same requirements as in Ref.~\cite{LHCb-PAPER-2016-030}, with the number of candidates corresponding to the size of the relevant background in the \LbToppipipi sample. Again, no $p$-values corresponding to a significance above 3 standard deviations are observed when the six configurations of the energy test probing \CP violation are applied to this sample. The background contribution from the \mbox{\decay{\Bz}{\Kp\pim\pip\pim}} decay is negligible within the mass window selected for the energy test. 

Finally, the proton detection asymmetry in simulation is replicated in the \mbox{\LbToppipipi} data sample by setting the \Lb flavour in the data sample at random to create the same asymmetry. The $P$-even and $P$-odd configurations of the energy test are then run for all three distance scales to test for effects that might lead to an incorrect rejection of the \CP-conserving hypothesis. 
This is repeated multiple times for each test with different flavour assignments for the \Lb candidates. In all six tests the distribution of $p$-values is consistent with being uniform, so no evidence for any bias from the proton detection asymmetry is found.

\label{sec:results}

The measured TPA from the fit to the full data set are $\aCPTodd=(-0.7\pm0.7\pm0.2)\%$ and $\aPTodd = (-4.0\pm0.7\pm0.2)\%$.
Consistency with the \CP-conserving hypothesis is observed, while a significant non-zero value for the \aPTodd asymmetry is found. The effect, estimated with the profile likelihood-ratio test, has a significance of $5.5$ standard deviations and indicates parity violation in the \LbToppipipi decay. 

\begin{figure}
    \centering
    \includegraphics[width=0.49\textwidth]{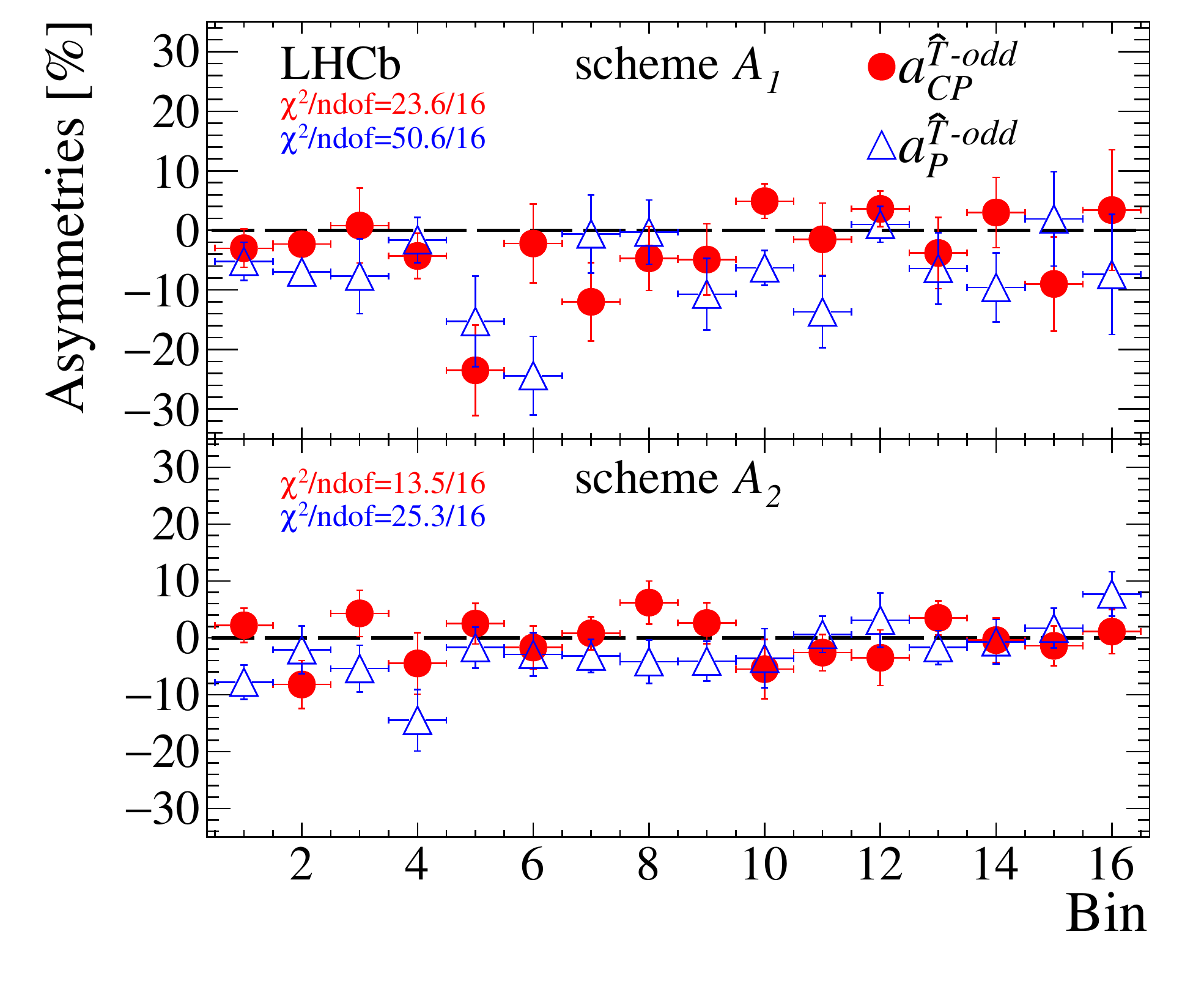}
    \includegraphics[width=0.49\textwidth]{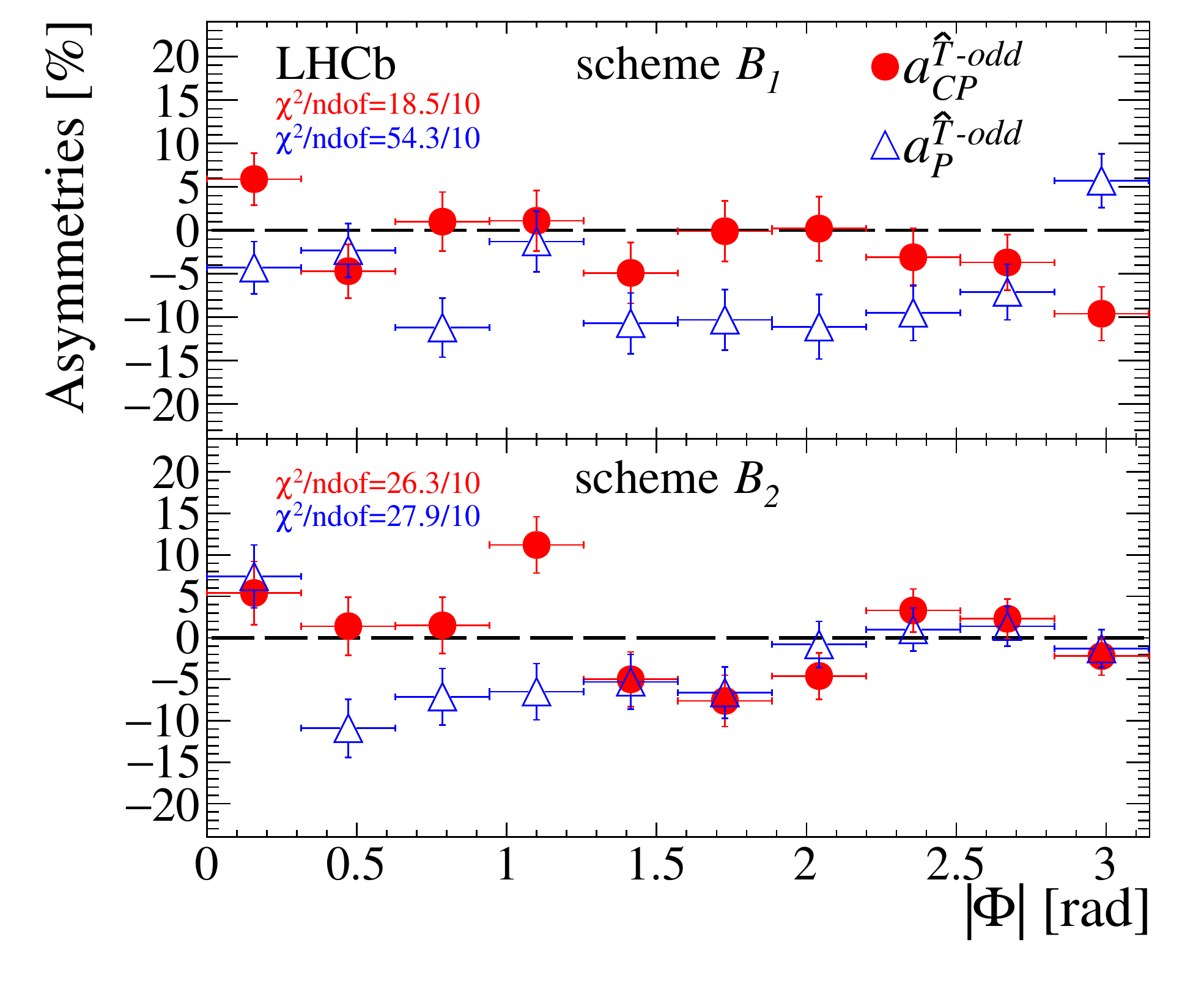}
    \caption{\small Measured asymmetries for the binning scheme (left) $A_1$ and $A_2$ and (right) $B_1$ and $B_2$. The error bars represent the sum in quadrature of the statistical and systematic uncertainties. The \chisq per ndof is calculated with respect to the null hypothesis and includes statistical and systematic uncertainties.}
    \label{fig:asymmetries TPA}
\end{figure}

The values of the TPA for the binning schemes $A_1$, $A_2$, $B_1$ and $B_2$ are shown in Fig.~\ref{fig:asymmetries TPA}.
In the binning schemes $A_2$ and $B_2$ the contribution from multiple $N^{*+}$ resonances dominates and therefore larger \CP asymmetries are possible relative to the  $A_1$ and $B_1$ binning schemes where the single $a_1$ resonance contributes. However, in the $A_2$ and $B_2$ phase-space regions, $p$-values with respect to the \CP-conserving hypothesis corresponding to statistical significances of $0.5$ and $2.9$ standard deviations are measured, respectively. The evidence of \CP violation previously observed~\cite{LHCb-PAPER-2016-030} is therefore not established.

The binning scheme $B$, which does not separate the $a_1$ and the $N^{*+}$ contributions, provides a deviation at 2.8 and 5.1 standard deviations from the \CP and $P$ conserving hypothesis, respectively. The compatibility of these results with the previous published measurements~\cite{LHCb-PAPER-2016-030}, based on the same binning scheme, is determined to be at 2.6 standard deviations, a value which decreases to 2.1 when the same BDT selection is applied.
Pseudoexperiments are generated by randomly assigning the flavour and \CT sign to each candidate. The asymmetries are extracted and the difference between the Run 1 and full datasets is determined as a \chisq value. The fraction of pseudoexperiments with a \chisq value greater than the observed \chisq in data represents the $p$-value.

The $p$-values measured in the case of binning schemes $A_1$ and $B_1$ indicate that the $P$ violation has a large contribution from the $\decay{\Lb}{\proton a_1(1260)^-}$ decay, for which the statistical significance is $5.5$ standard deviations. 

The $p$-values obtained for different configurations of the energy test are summarised in Table~\ref{Unblinded}. All \CP-violation searches using the energy test result in $p$-values with a significance of 3 standard deviations or smaller. 
Given the reported $p$-value for the \mbox{$P$-even} configuration of the energy test at a distance scale of 2.7 GeV$^{2}$/$c^4$ is marginally consistent with the \CP-conserving hypothesis, the different distance scales considered are combined to obtain a global $p$-value for the \mbox{$P$-even} configuration. A new test statistic is defined as $Q=p_1 \, p_2 \, p_3$, where $p_i$ corresponds to a $p$-value for a distance scale $i$. The value of $Q$ observed in data is then compared to the corresponding values from permutations, considering correlations between the different distance scales. The combined $p$-value for the $P$-even energy test configuration is $4.6\times10^{-3}$.
In addition, the test for parity violation is also performed using the same three distance scales with the energy test. The results are reported in Table~\ref{Unblinded}. The $p$-values found with this study correspond to the observation of local parity violation for the two smaller distance scales probed with the highest significance observed to be $5.3$ standard deviations. 

\begin{table}[tb]
\centering
\caption{\label{Unblinded} The $p$-values from the energy test for different distance scales and test configurations.}
\label{unblindedtab}
\begin{tabular}{l|ccc}
Distance scale $\delta$       & 1.6 GeV$^2/c^4$  & 2.7 GeV$^2/c^4$ & 13 GeV$^2/c^4$   \\\hline
$p$-value (\CP conservation, $P$ even) & $3.1 \times 10^{-2}$ & $2.7 \times 10^{-3}$ & $1.3 \times 10^{-2}$\\
$p$-value (\CP conservation, $P$ odd) & $1.5 \times 10^{-1}$ & $6.9 \times 10^{-2}$ & $6.5 \times 10^{-2}$   \\
$p$-value ($P$ conservation) & $1.3 \times 10^{-7}$ & $4.0 \times 10^{-7}$ & $1.6 \times 10^{-1}$   \\
\end{tabular}
\end{table}

\label{sec:conclusions}
In conclusion, this Letter reports the searches for \CP violation in \LbToppipipi decays both globally and in regions of phase space, using two different methods. The results are marginally compatible with the no \CP-violation hypothesis. 
Violation of $P$ symmetry is observed using both methods, locally with a significance of over 5 standard deviations, and, when the triple product asymmetries are evaluated having integrated over the entire sample, with a significance of $5.5$ standard deviations. 


\section*{Acknowledgements}
%
%
\noindent We express our gratitude to our colleagues in the CERN
accelerator departments for the excellent performance of the LHC. We
thank the technical and administrative staff at the LHCb
institutes.
We acknowledge support from CERN and from the national agencies:
CAPES, CNPq, FAPERJ and FINEP (Brazil); 
MOST and NSFC (China); 
CNRS/IN2P3 (France); 
BMBF, DFG and MPG (Germany); 
INFN (Italy); 
NWO (Netherlands); 
MNiSW and NCN (Poland); 
MEN/IFA (Romania); 
MSHE (Russia); 
MinECo (Spain); 
SNSF and SER (Switzerland); 
NASU (Ukraine); 
STFC (United Kingdom); 
DOE NP and NSF (USA).
We acknowledge the computing resources that are provided by CERN, IN2P3
(France), KIT and DESY (Germany), INFN (Italy), SURF (Netherlands),
PIC (Spain), GridPP (United Kingdom), RRCKI and Yandex
LLC (Russia), CSCS (Switzerland), IFIN-HH (Romania), CBPF (Brazil),
PL-GRID (Poland) and OSC (USA).
We are indebted to the communities behind the multiple open-source
software packages on which we depend.
Individual groups or members have received support from
AvH Foundation (Germany);
EPLANET, Marie Sk\l{}odowska-Curie Actions and ERC (European Union);
ANR, Labex P2IO and OCEVU, and R\'{e}gion Auvergne-Rh\^{o}ne-Alpes (France);
Key Research Program of Frontier Sciences of CAS, CAS PIFI, and the Thousand Talents Program (China);
RFBR, RSF and Yandex LLC (Russia);
GVA, XuntaGal and GENCAT (Spain);
the Royal Society
and the Leverhulme Trust (United Kingdom).

\appendix

\section{Definition of the binning scheme A}
\label{app:definition binning A}

The definition of the binning scheme $A$ is reported in Table~\ref{tab:new binning scheme}. 

\begin{table}[htb]
\centering
\caption{\small Definition of binning scheme $A$. This binning scheme is based on the helicity angles of the decay topology $\decay{\Lb}{\left(\decay{N^{*+}}{\left(\decay{\Deltares^{++}}{\proton\pip}\right)\pim}\right)\pim}$ where $\varphi$ is the azimuthal angle of the proton in the $\Delta^{++}$ rest frame and $\theta_{\Deltares^{++}}$ ($\theta_\proton$) is the polar angle of the $\Deltares^{++}$ ($\proton$) in the $N^{*+}$ ($\Deltares^{++}$) rest frame.}
\label{tab:new binning scheme}
\resizebox{\linewidth}{!}{\begin{tabular}{ccc}
\hline
Bin number & Polar angles & Azimuthal angles\\
\hline
\multirow{2}{*}{1} & $\theta_p \in [0,\pi/4] \text{ and } \theta_{\Delta^{++}} \in [0,\pi/4]$ & \multirow{2}{*}{$\lvert\varphi\rvert \in [0,\pi/2]$}\\
& $\theta_p \in [\pi/2,3\pi/4] \text{ and } \theta_{\Delta^{++}} \in [\pi/2,3\pi/4]$ &\\ \hline

\multirow{2}{*}{2} & $\theta_p \in [0,\pi/4] \text{ and } \theta_{\Delta^{++}} \in [\pi/4,\pi/2]$ & \multirow{2}{*}{$\lvert\varphi\rvert \in [0,\pi/2]$}\\
& $\theta_p \in [\pi/2,3\pi/4] \text{ and } \theta_{\Delta^{++}} \in [3\pi/4,\pi]$ &\\ \hline

\multirow{2}{*}{3} & $\theta_p \in [0,\pi/4] \text{ and } \theta_{\Delta^{++}} \in [\pi/2,3\pi/4]$ &  \multirow{2}{*}{$\lvert\varphi\rvert \in [0,\pi/2]$}\\
& $\theta_p \in [\pi/2,3\pi/4] \text{ and } \theta_{\Delta^{++}} \in [0,\pi/4]$ &\\ \hline

\multirow{2}{*}{4} & $\theta_p \in [0,\pi/4] \text{ and } \theta_{\Delta^{++}} \in [3\pi/4,\pi]$ &  \multirow{2}{*}{$\lvert\varphi\rvert \in [0,\pi/2]$}\\
& $\theta_p \in [\pi/2,3\pi/4] \text{ and } \theta_{\Delta^{++}} \in [\pi/4,\pi/2]$ &\\ \hline

\multirow{2}{*}{5} & $\theta_p \in [\pi/4,\pi/2] \text{ and } \theta_{\Delta^{++}} \in [0,\pi/4]$ & \multirow{2}{*}{$\lvert\varphi\rvert \in [0,\pi/2]$}\\
& $\theta_p \in [3\pi/4,\pi] \text{ and } \theta_{\Delta^{++}} \in [\pi/2,3\pi/4]$ &\\ \hline

\multirow{2}{*}{6} & $\theta_p \in [\pi/4,\pi/2] \text{ and } \theta_{\Delta^{++}} \in [\pi/4,\pi/2]$ & \multirow{2}{*}{$\lvert\varphi\rvert \in [0,\pi/2]$}\\
& $\theta_p \in [3\pi/4,\pi] \text{ and } \theta_{\Delta{++}} \in [3\pi/4,\pi]$ &\\ \hline

\multirow{2}{*}{7} & $\theta_p \in [\pi/4,\pi/2] \text{ and } \theta_{\Delta^{++}} \in [\pi/2,3\pi/4]$ & \multirow{2}{*}{$\lvert\varphi\rvert \in [0,\pi/2]$}\\
& $\theta_p \in [3\pi/4,\pi] \text{ and } \theta_{\Delta^{++}} \in [0,\pi/4]$ &\\ \hline

\multirow{2}{*}{8} & $\theta_p \in [\pi/4,\pi/2] \text{ and } \theta_{\Delta^{++}} \in [3\pi/4,\pi]$ & \multirow{2}{*}{$\lvert\varphi\rvert \in [0,\pi/2]$}\\
& $\theta_p \in [3\pi/4,\pi] \text{ and } \theta_{\Delta^{++}} \in [\pi/4,\pi/2]$ &\\ \hline

\multirow{2}{*}{9} & $\theta_p \in [0,\pi/4] \text{ and } \theta_{\Delta^{++}} \in [0,\pi/4]$ & \multirow{2}{*}{$\lvert\varphi\rvert \in [\pi/2,\pi]$}\\
& $\theta_p \in [\pi/2,3\pi/4] \text{ and } \theta_{\Delta^{++}} \in [\pi/2,3\pi/4]$ &\\ \hline

\multirow{2}{*}{10} & $\theta_p \in [0,\pi/4] \text{ and } \theta_{\Delta^{++}} \in [\pi/4,\pi/2]$ & \multirow{2}{*}{$\lvert\varphi\rvert \in [\pi/2,\pi]$}\\
& $\theta_p \in [\pi/2,3\pi/4] \text{ and } \theta_{\Delta^{++}} \in [3\pi/4,\pi]$ &\\ \hline

\multirow{2}{*}{11} & $\theta_p \in [0,\pi/4] \text{ and } \theta_{\Delta^{++}} \in [\pi/2,3\pi/4]$ & \multirow{2}{*}{$\lvert\varphi\rvert \in [\pi/2,\pi]$}\\
& $\theta_p \in [\pi/2,3\pi/4] \text{ and } \theta_{\Delta^{++}} \in [0,\pi/4]$ &\\ \hline

\multirow{2}{*}{12} & $\theta_p \in [0,\pi/4] \text{ and } \theta_{\Delta^{++}} \in [3\pi/4,\pi]$ & \multirow{2}{*}{$\lvert\varphi\rvert \in [\pi/2,\pi]$}\\
& $\theta_p \in [\pi/2,3\pi/4] \text{ and } \theta_{\Delta^{++}} \in [\pi/4,\pi/2]$ &\\ \hline

\multirow{2}{*}{13} & $\theta_p \in [\pi/4,\pi/2] \text{ and } \theta_{\Delta^{++}} \in [0,\pi/4]$ & \multirow{2}{*}{$\lvert\varphi\rvert \in [\pi/2,\pi]$}\\
& $\theta_p \in [3\pi/4,\pi] \text{ and } \theta_{\Delta^{++}} \in [\pi/2,3\pi/4]$ &\\ \hline

\multirow{2}{*}{14} & $\theta_p \in [\pi/4,\pi/2] \text{ and } \theta_{\Delta^{++}} \in [\pi/4,\pi/2]$ & \multirow{2}{*}{$\lvert\varphi\rvert \in [\pi/2,\pi]$}\\
& $\theta_p \in [3\pi/4,\pi] \text{ and } \theta_{\Delta^{++}} \in [3\pi/4,\pi]$ &\\ \hline

\multirow{2}{*}{15} & $\theta_p \in [\pi/4,\pi/2] \text{ and } \theta_{\Delta^{++}} \in [\pi/2,3\pi/4]$ & \multirow{2}{*}{$\lvert\varphi\rvert \in [\pi/2,\pi]$}\\
& $\theta_p \in [3\pi/4,\pi] \text{ and } \theta_{\Delta^{++}} \in [0,\pi/4]$ &\\ \hline

\multirow{2}{*}{16} & $\theta_p \in [\pi/4,\pi/2] \text{ and } \theta_{\Delta^{++}} \in [3\pi/4,\pi]$ & \multirow{2}{*}{$\lvert\varphi\rvert \in [\pi/2,\pi]$}\\
& $\theta_p \in [3\pi/4,\pi] \text{ and } \theta_{\Delta^{++}} \in [\pi/4,\pi/2]$ &\\
\hline
\end{tabular}
}
\end{table}

\FloatBarrier
\addcontentsline{toc}{section}{References}
\bibliographystyle{LHCb}
\bibliography{main,standard,LHCb-PAPER,LHCb-CONF,LHCb-DP,LHCb-TDR}

\newpage

\centerline
{\large\bf LHCb collaboration}
\begin
{flushleft}
\small
R.~Aaij$^{31}$,
C.~Abell{\'a}n~Beteta$^{49}$,
T.~Ackernley$^{59}$,
B.~Adeva$^{45}$,
M.~Adinolfi$^{53}$,
H.~Afsharnia$^{9}$,
C.A.~Aidala$^{79}$,
S.~Aiola$^{25}$,
Z.~Ajaltouni$^{9}$,
S.~Akar$^{64}$
% ...

{\footnotesize \it

% ...
$ ^{9}$Universit{\'e} Clermont Auvergne, CNRS/IN2P3, LPC, Clermont-Ferrand, France\\
% ...
$ ^{25}$INFN Sezione di Milano, Milano, Italy\\
% ...
$ ^{31}$Nikhef National Institute for Subatomic Physics, Amsterdam, Netherlands\\
% ...
$ ^{45}$Instituto Galego de F{\'\i}sica de Altas Enerx{\'\i}as (IGFAE), Universidade de Santiago de Compostela, Santiago de Compostela, Spain\\
% ...
$ ^{49}$Physik-Institut, Universit{\"a}t Z{\"u}rich, Z{\"u}rich, Switzerland\\
% ...
$ ^{53}$H.H. Wills Physics Laboratory, University of Bristol, Bristol, United Kingdom\\
% ...
$ ^{59}$Oliver Lodge Laboratory, University of Liverpool, Liverpool, United Kingdom\\
% ...
$ ^{64}$University of Cincinnati, Cincinnati, OH, United States\\
% ...
$ ^{67}$Syracuse University, Syracuse, NY, United States\\
% ...
$ ^{79}$University of Michigan, Ann Arbor, United States, associated to $^{67}$\\
\bigskip
% ...
}
\end{flushleft}

\end{document}